\documentclass[conf, onecolumn]{ao4elt3}

\pdfoutput=1

\usepackage{graphics}  %% Note: this is not required by the style, you may 
\usepackage{graphicx}  %% Note: this is not required by the style, you may 

\usepackage[varg]{txfonts} % Times fonts
\usepackage[latin1]{inputenc}
\usepackage{color}
\usepackage{soul}

\begin{document}
\pagestyle{plain}
\title{GeMS FIRST SCIENCE RESULTS}
\author{Benoit Neichel\inst{1,2}\thanks{benoit.neichel@lam.fr} \and Fabrice Vidal\inst{1} \and Francois Rigaut\inst{3}  \and Eleazar Rodrigo Carrasco\inst{1} \and Gustavo Arriagada\inst{1} \and Andrew Serio\inst{1} \and Peter Pessev\inst{1} \and Claudia Winge\inst{1} \and Marcos van Dam\inst{1,4} \and Vincent Garrel\inst{1}  \and Constanza Araujo\inst{1} \and Maxime Boccas\inst{1} \and Vincent Fesquet\inst{1} \and Ramon Galvez\inst{1} \and  Gaston Gausachs\inst{1} \and Javier L\"{u}hrs\inst{1} \and Vanessa Montes\inst{1} \and Cristian Moreno\inst{1} \and William Rambold\inst{1} \and Chadwick Trujillo\inst{1} \and Cristian Urrutia\inst{1} \and Tomislav Vucina\inst{1}
}

\institute{Gemini Observatory, c/o AURA, Casilla 603, La Serena, Chile \and Aix Marseille Universit\'{e}, CNRS, LAM (Laboratoire d'Astrophysique de Marseille) UMR 7326, 13388, Marseille, France \and The Australian National University, RSAA, Mount Stromlo Observatory, Cotter Road, Weston Creek ACT 2611, Australia \and Flat Wavefronts, PO BOX 1060, Christchurch 8140, New Zealand}
\abstract{
After 101 nights of commissioning, the Gemini MCAO system (GeMS) started science operations in December 2012.
% I thought a 101 is better than 100. It sounds like "The 1001 nights" !!!
After a brief reminder on GeMS specificities, we describe the overall GeMS performance, and we focus then on the first science results obtained with GeMS, illustrating the unique capabilities of this new Gemini instrument.} %end of abstract
\maketitle
\section{Brief introduction to GeMS}
\label{intro}
\subsection{GeMS overview}
The Gemini Multi-conjugate adaptive optics System (GeMS) at the Gemini South telescope in Cerro Pachon is the first and only sodium-based multi-Laser Guide Star (LGS) adaptive optics system. It has been described in details in previous papers \cite{rigaut2013review,neichel2013review,rigaut2012gems,dorgeville2012gemini,neichel2012science,fesquet2013lgsf}, and here we only briefly recall its main characteristics. GeMS works with a LGS constellation of 5-spots: 4 of the LGS spots are at the corners of a 60 arcsec square, with the 5th positioned in the center. These LGSs are produced by a 50W laser split into 5 distinct 10-Watt beacons by a series of beamsplitters. The laser bench and its electronics enclosure are housed inside a Laser Service Enclosure (LSE), an 8x2m clean room seating on an extension of the telescope elevation platform (a Nasmyth focus for other telescopes). The laser beam is relayed from the LSE to the Laser Launch Telescope (LLT) located behind the secondary mirror of the Gemini telescope by a set of mirrors called BTO for Beam Transfer Optics. The BTO also allows for the LGS constellation alignment, beam shuttering and laser beam quality monitoring. The AO bench (called Canopus) is mounted on one of the Cassegrain port. The F/16 Gemini beam is directed toward Canopus by the AO fold. The MCAO correction is then performed by two Deformable Mirrors (DMs) conjugated to 0 and 9 km and one Tip-Tilt Mirror (TTM). After this, a first dichroic beam splitter is responsible for separating the visible and NIR light, sending the former to the WaveFront Sensors (WFS) and the latter to the science output with a F/33.2 focal-ratio to feed the instruments. GeMS is a facility instrument, and as such it can direct its light output to different science instruments installed at the Cassegrain focus of the Gemini South telescope. The main instrument used to date is GSAOI \cite{carrasco2012results}, a ``wide-field'' ($85''\times85''$) camera designed to work at the diffraction limit of the 8-meter telescope in the near-infrared (NIR). 

\subsection{GeMS performance}\label{sec:1}
A very detailed analysis of the GeMS performance is given in a companion paper ~\cite{Vidal2013}. Here we summarize the overall delivered image quality. The GeMS Strehl Ratios (SR) and Full Width Half Max (FWHM) performance presented in this section are based on data collected over 33 nights over the December 2012 to June 2013 period. The delivered Strehl ratios and FWHM measured under different natural seeing conditions are shown in Table \ref{tab:gems_performance1}. The results are based on images observed with a constellation of 3 NGS and with exposure times between 10 and 180 seconds. Lower performance can be expected if the observations use less than three NGS.

\begin{table}[h]
\caption{GeMS overall performance.}
\begin{center}
\begin{tabular}{l|ccc|ccc}
\hline \hline
%                 & J & H & K \\ 
% Seeing at 550nm & FWHM - SR & FWHM - SR & FWHM - SR \\ \hline
% $<$0.45"     & 80 mas - 10\% & 70 mas - 15\% & 60 mas - 30\% \\
% 0.45" - 0.8"      & 130 mas - 5\% & 100 mas - 10\% & 90 mas - 15\% \\
% 0.8" - 1.00"      & 150 mas - 2\% & 130 mas - 5\% & 120 mas - 10\% \\ \hline \hline
             & \multicolumn{3}{|c|}{FWHM [mas]} & \multicolumn{3}{|c}{Strehl ratio [\%]} \\
Seeing at 550nm & J & H & K & J & H & K \\ \hline
$<$0.45$''$     &  80  &  70  &  60  & 10  & 15  & 30 \\
0.45$''$ - 0.8$''$ & 130  & 100  &  90  &  5  & 10  & 15 \\
0.8$''$ - 1.00$''$ & 150  & 130  & 120  &  2  &  5  & 10 \\ \hline \hline
\end{tabular}
\end{center}
\label{tab:gems_performance1}
\end{table}

The delivered FWHM and Strehl ratios not only depend on the seeing and the selected NGS constellation, the number of NGSs and their brightnesses, but also depend on the LGS photons return (this parameter varies seasonally), turbulence speed ($\tau_0$) and profile (Cn2), non-common path aberrations and other AO optimization and calibration parameters. Hence, another way to present the performance is to look at the delivered image quality for a given fraction of the observing time. This is what is presented in Table \ref{tab:gems_performance2}. For instance, we see that 50\% of the time, GeMS delivers a FWHM of 75mas (or better) in H-band.

\begin{table}[h]
\caption{GeMS overall performance, fractional view}
\begin{center}
% \begin{tabular}{llll}
% \hline \hline
%        & 20 percentile & 50 percentile & 70 percentile \\
%        & FWHM - SR & FWHM - SR & FWHM - SR \\ \hline
% K-band    & 76 mas - 26\% & 95 mas - 17\%& 110 mas -13\% \\
% H-band    & 64 mas - 15\% & 75 mas - 11\% & 90 mas - 8\%\\
% J-band     & 64 mas - 8\% & 87 mas - 5\% & 110 mas - 3\% \\ \hline \hline
% \end{tabular}
\begin{tabular}{l|ccc|ccc} \hline \hline
      & \multicolumn{3}{|c|}{FWHM [mas]} & \multicolumn{3}{|c}{Strehl ratio [\%]} \\
Seeing conditions &    J &  H &   K &  J &  H &  K \\ \hline
20 percentile   &   64 & 64 &  76 &  8 & 15 & 26 \\
50 percentile   &   87 & 75 &  95 &  5 & 11 & 17 \\
70 percentile   &  110 & 90 & 110 &  3 &  8 & 13 \\ \hline \hline
\end{tabular}
\end{center}
\label{tab:gems_performance2}
\end{table}

It is important to note that the current performance of GeMS is under the original specification, roughly by a factor of 2 in terms of SR. There are several reasons for this, the main two being that:
\begin{itemize}
\item GeMS was designed and integrated with 3 DMs. However, following issues with one of them, the system has been working with only 2 DMs, the ground and the 9km DMs. This reduced the number of active actuators from 684 (design) down to 360 (current). The current system is then much more sensitive to changes in the Cn2 profile, and generalized fitting (tomographic error) is significant. The current two DMs system can be viewed as a fairly potent GLAO system (17x17 actuators across M1) with an additional low order DM at altitude (9x9 actuator across M1).
\item We generally work at about 140 to 160ph/subaperture/frame or between 35 and 40ph/pixel/frame, as the LGSWFS is working with 2x2pixels/subaperture. To maintain this flux level, the sampling frequency is adjusted: During the low sodium season (December and January) a typical guide rate is about 200 Hz, while during high sodium season (April and May) the guide rate varies between 400-800 Hz. The system was designed and can work up to 800Hz, hence when at 200Hz, the servolag \& noise errors can be very large.
\end{itemize}

\subsection{GeMS System Verification}
The System Verification (SV) observations started in December 2012, one year after the instrument first light, and lasted 3 months. The SV observations provide an end-to-end test of a new instrument or capability, from the proposal process to data delivery, prior to offering it to the community for general use. With GeMS/GSAOI, one main objective was to demonstrate the gain brought by MCAO on a large variety of science topics, including extended sources, crowded fields, and faint targets. Twenty-three programs were submitted for a total of 138 hours, of which 13 were selected for execution between December 2012 and March 2013, for a total of 60 hours. 12 targets out of the 13 selected were observed during the course of 18 nights. The system efficiency shows that about 20\% of extra time was required to complete the programs, and about 20\% of the observing time was lost due to fault. The efficiency was improved during the first semester of science operations. In the next section we describe some of the data acquired during the SV period, and the first science results obtained. Note that two science papers have already been published based on GeMS/GSAOI SV data at the time of this writing (\cite{Davidge,Ronald}).

\section{Science with MCAO}

As stated above, the GeMS/GSAOI SV period was used to test the instrument under different conditions, and for different kind of science programs. There are several reasons to use an MCAO system. Some of them are: the wide field capability, the sky coverage and the good astrometric performance. 
% \footnote{They are of course, many other valid reasons that could lead someone to use and MCAO system.} 

\subsection{MCAO and the wide field of view}

Unsurprisingly, the first reason why astronomers use MCAO is because of the wide, corrected, and uniform Field of View (FoV). Below, we show some examples of programs that benefit from a large FoV.

\begin{figure}[!ht]
\includegraphics[width = 0.6\linewidth]{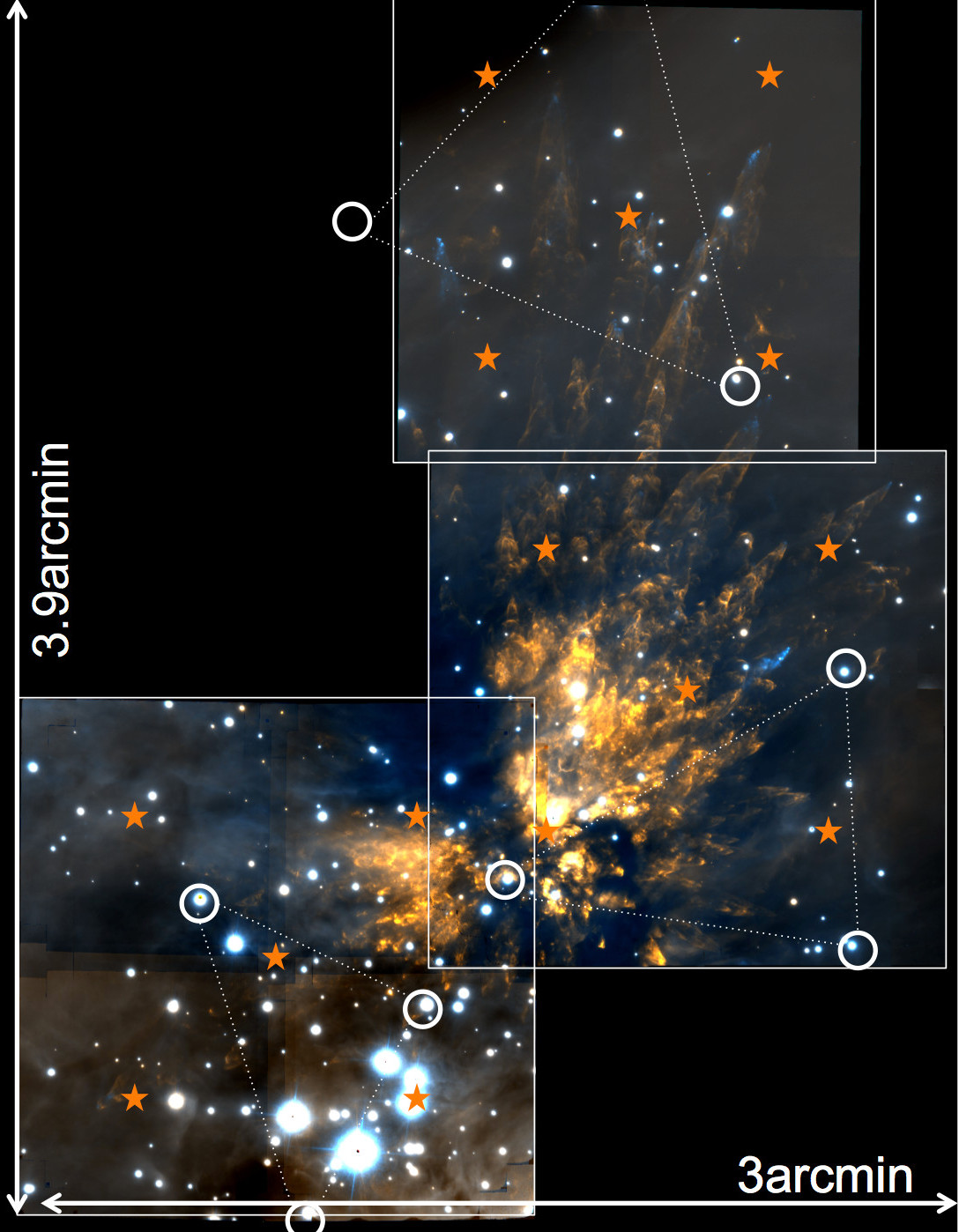} 
\includegraphics[width = 0.4\linewidth]{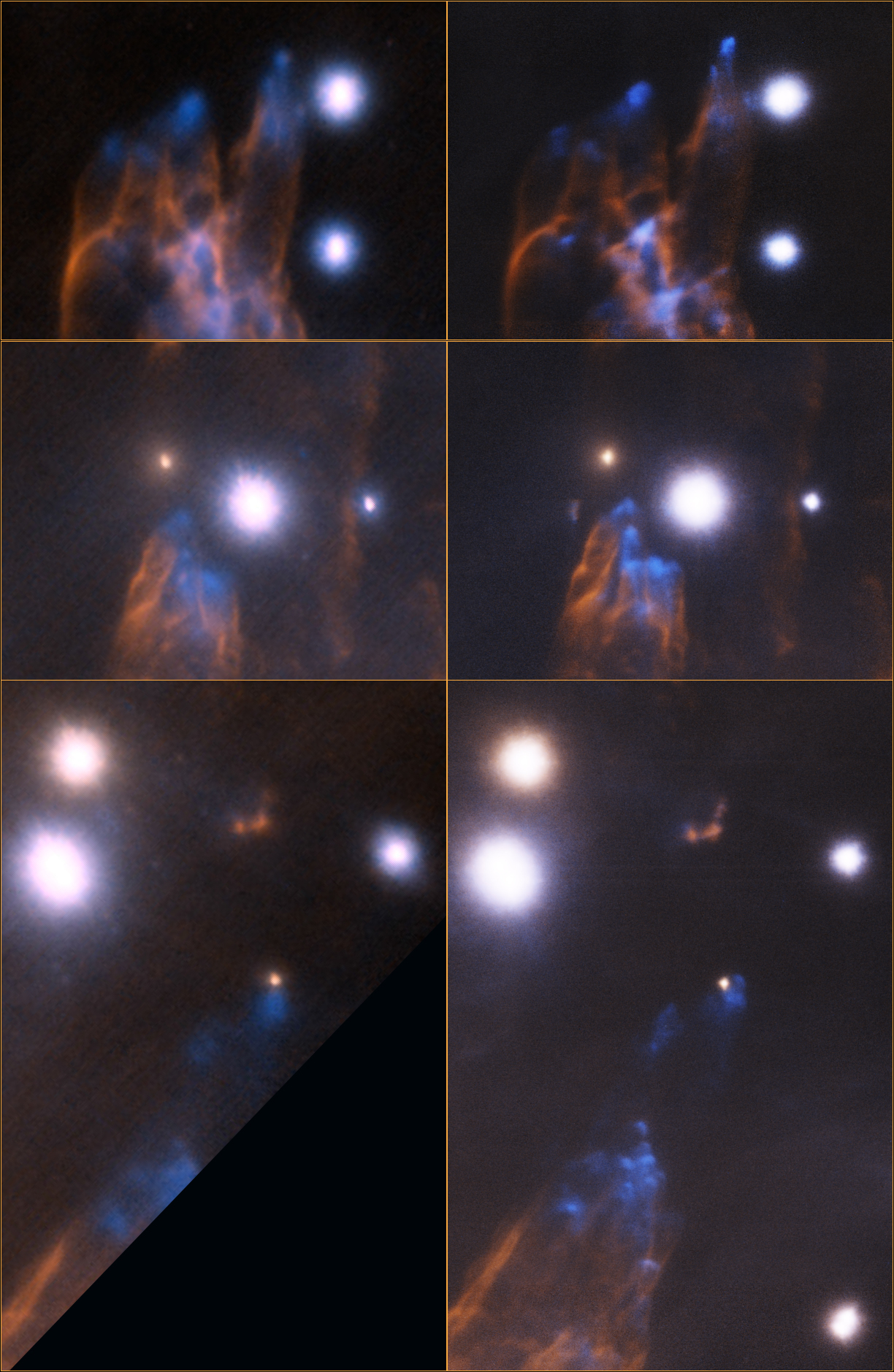}
\caption{{\bf Left:} Orion Bullets - SV program led by John Bally. Mosaic of 3 GeMS/GSAOI pointings made from FeII, H2, and K-2.2 microns filters (blue, orange, and white, respectively). The field-of-view is 3.0 x 3.9 arcminutes and is oriented with the north up. The total (integrated) exposure time was 30 minutes cumulative for all filters and fields. The white circles show the location of the NGSs used for Tip-Tilt correction. The orange symbol stars in the image indicate the position of the LGS. {\bf Right:} Detailed views of the Orion Bullet region. In each image pair, left is the Altair 2007 image and right is the new 2012 GeMS image. By looking carefully, one can actually see that the bullets (blue knots) have moved over the years. Image Credit: Gemini Observatory/AURA.}
\label{fig:1}       % Give a unique label
\end{figure}

% there is no "noindent" in ao4elt3_template.tex
% \noindent
The first target that was observed during the SV period was the Orion Bullets (see Fig. \ref{fig:1}). The ``Bullets" region of the Orion nebula has a long history of AO imaging at Gemini: A %in English, capitalize after a colon (most of the time): http://grammar.ccc.commnet.edu/grammar/marks/colon.htm
smaller section of the field shown here was first targeted with the Altair AO system at Gemini North in 2007 (see right-hand part of Fig. \ref{fig:1}). The features in the Orion Nebula are clumps of gas violently ejected from an unknown event associated with the recent formation of a cluster of massive stars. The strong winds produced by this ``explosion" expelled these bullets of gas at supersonic speeds, leaving behind the distinctive tubular and cone-shaped wakes. The PIs of these observations (John Bally and Adam Ginsburg from the University of Colorado) want to compare high-angular-resolution images of this region over several years (including the Altair images obtained in 2007) in order to measure the bullets motions. By mapping the proper motions of the bullets, they can build a complete 3D dynamical model of the region, and try to determine the intensity of the blast and the nature of the bullets (e.g. \cite{Bally}). The sub-arcsecond resolution provided by GeMS is needed to resolve the shocks and to search for the compact, high-density bullets, responsible for these wakes.

This Orion image also illustrates the gain brought by MCAO. The final mosaic image, made by three GeMS/GSAOI pointings, covers a field-of-view  of 4 x 3 arcminutes, resulting in one of the largest in size AO-corrected image ever obtained. This is the main advantage of MCAO when compared to other AO systems: In one shot, the area of sky covered is 10- to 20-times larger than any previous AO system, and the performance remains uniform over this large field. As an illustration, the right-hand part of Fig.\ref{fig:1} shows a direct comparison of the 2007 Altair's image, with the new GeMS/GSAOI ones. As Altair is a classical AO system, the corner of the image shows elongation due to anisoplanatism. GeMS, with the DM conjugated in altitude is able to compensate some of the turbulence located at high altitudes, and delivers a uniform correction across the entire field.

\begin{figure}[!ht]
\centerline{\resizebox{0.9\columnwidth}{!}{\includegraphics{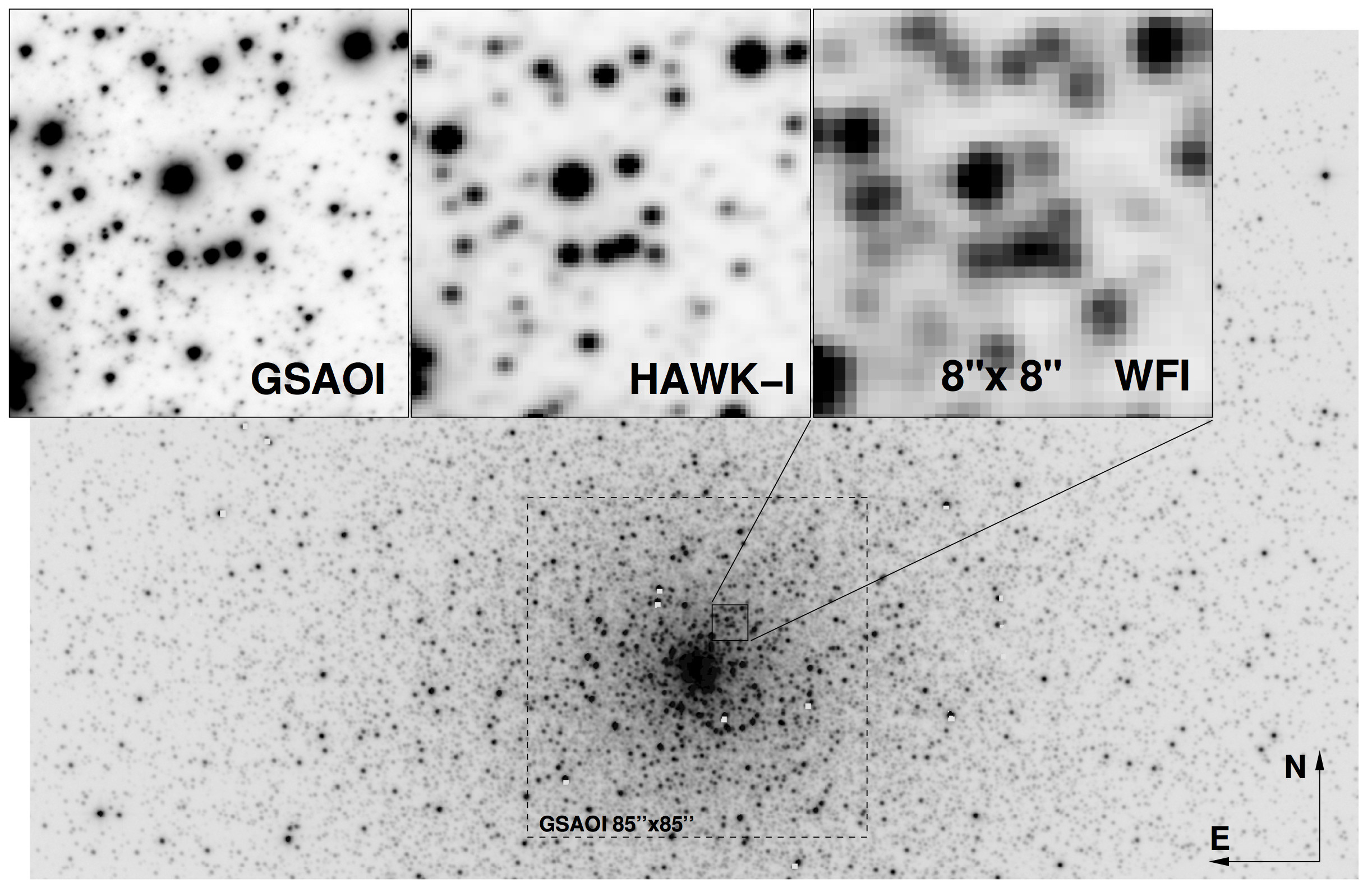} }}
\caption{NGC 1851 - SV program led by Alan McConnachie. Image Courtesy of Mischa Schirmer. NGC1851 is an ancient globular star cluster some 40,000 light-years from our Sun. The three inserts show a comparison of the resolution delivered by different instrument: GeMS/GSAOI, Hawk-I (a NIR seeing-limited imager mounted at the ESO-VLT) and WFI (a NIR seeing-limited imager installed at the 2.2m MPG/ESO telescope in La Silla). The angular resolution is 0.08 arc seconds on the GeMS image, 0.33 arc seconds in the Hawk-I image, and 0.75 arc seconds in the WFI one.}
\label{fig:3}       % Give a unique label
\end{figure}

Fig.\ref{fig:3} shows another example of an image acquired during GeMS/GSAOI SV. The target is a globular cluster, NGC1851 (\cite{Fiorentino}). Star clusters are one of the main science cases for MCAO. Crowded fields are where AO brings its largest gains. By ``compacting'' the Point Spread Function it bring out the faintest stars in the cluster which are crucial to study the star formation in these environments.
% the AO correction ``deblend" multiple systems in crowded fields, allowing astronomers to access the cluster's fainter stars, which are crucial in studies of star formation in these different environments. 
% Moreover, 
In addition, by delivering a uniform performance over fields that encompass most globular star cluster sizes, MCAO greatly improves the photometric precision on these crowded fields, and opens the way for a better understanding of the cluster's stellar population, particularly of its age, any evidence for multiple stellar populations, and the distribution of low mass stars. As a demonstration of the huge potential of MCAO for the star cluster science case, one can look at the (at least) 11 papers published based on MAD data (ESO's MCAO demonstrator).

\begin{figure}[!ht]
\centerline{\resizebox{0.96\columnwidth}{!}{\includegraphics{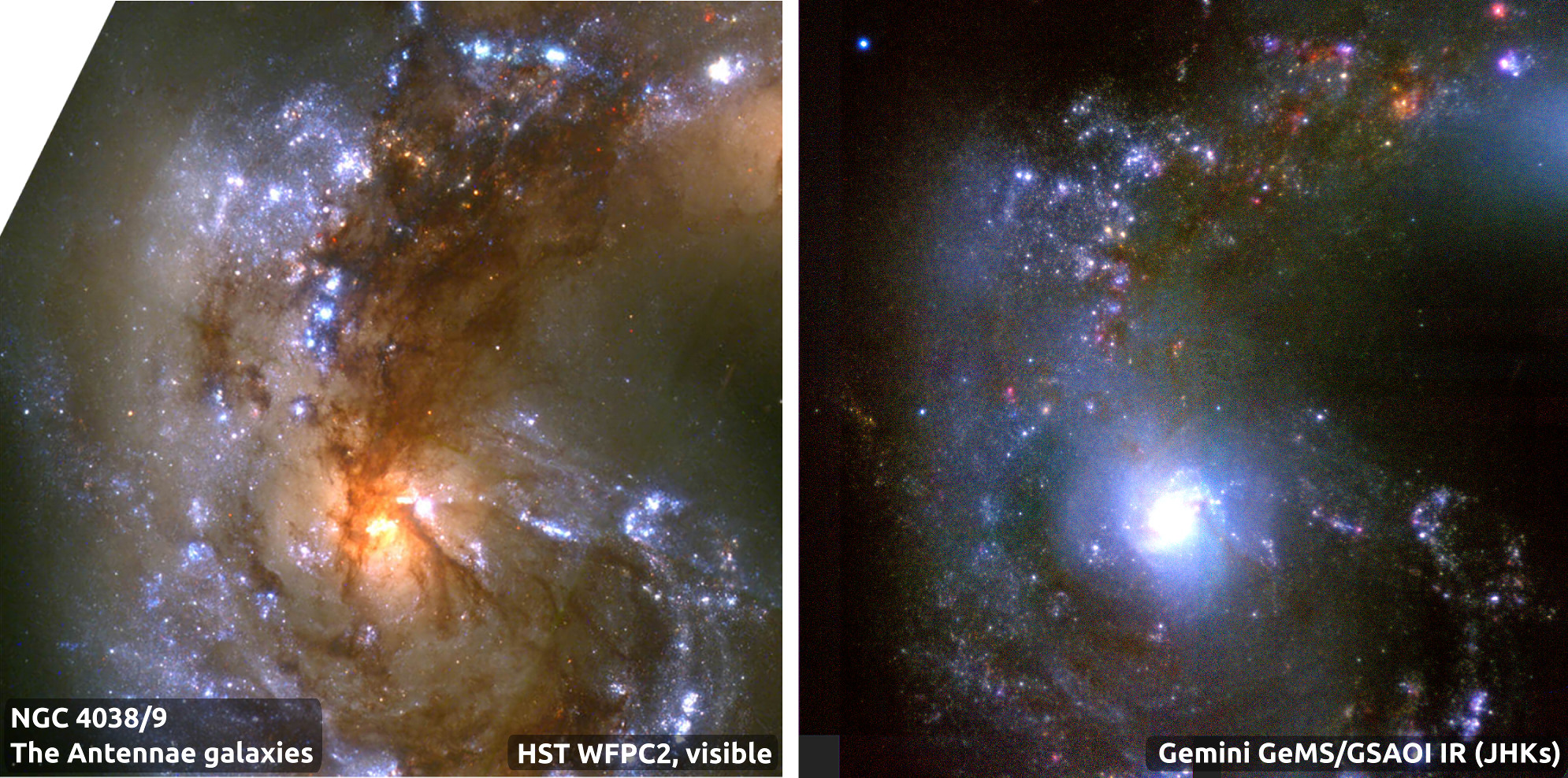} }}
\caption{The Antennae galaxy as seen by the Hubble Space Telescope (left, composite visible image) and GeMS (right, composite infrared image). %Commissioning target.
Because of the amount of dust, largely opaque to visible lights, the view offered by HST and by GeMS are significantly different. GeMS's infrared view, at an angular resolution similar or slightly better than HST in the visible, provides a useful complementary information to study these astrophysical objects.}
\label{fig:4}       % Give a unique label
\end{figure}

Finally, a third example to illustrate the gain brought by MCAO in terms of field of view is the Antennae galaxies shown in Fig. \ref{fig:4}. This is probably the most recognized pair of interacting disk galaxies in the sky. The popular name comes from the resemblance of their tidal tails to the antennae of an insect, as seen in the wide-field images. The starburst system, only about 10.5 million light-years distant, harbors a rich population of massive young clusters, whose formation has been triggered by the interaction. Considered to be globular cluster progenitors, these objects are resolved in the GeMS image of NGC 4038. The high resolution data provided by GeMS/GSAOI allows to differentiate compact star clusters from individual stars, study their integrated-light properties, and set constraints on the underlying stellar populations.

\subsection{MCAO and the sky coverage}

\begin{figure}[!ht]
\centerline{\resizebox{0.9\columnwidth}{!}{\includegraphics{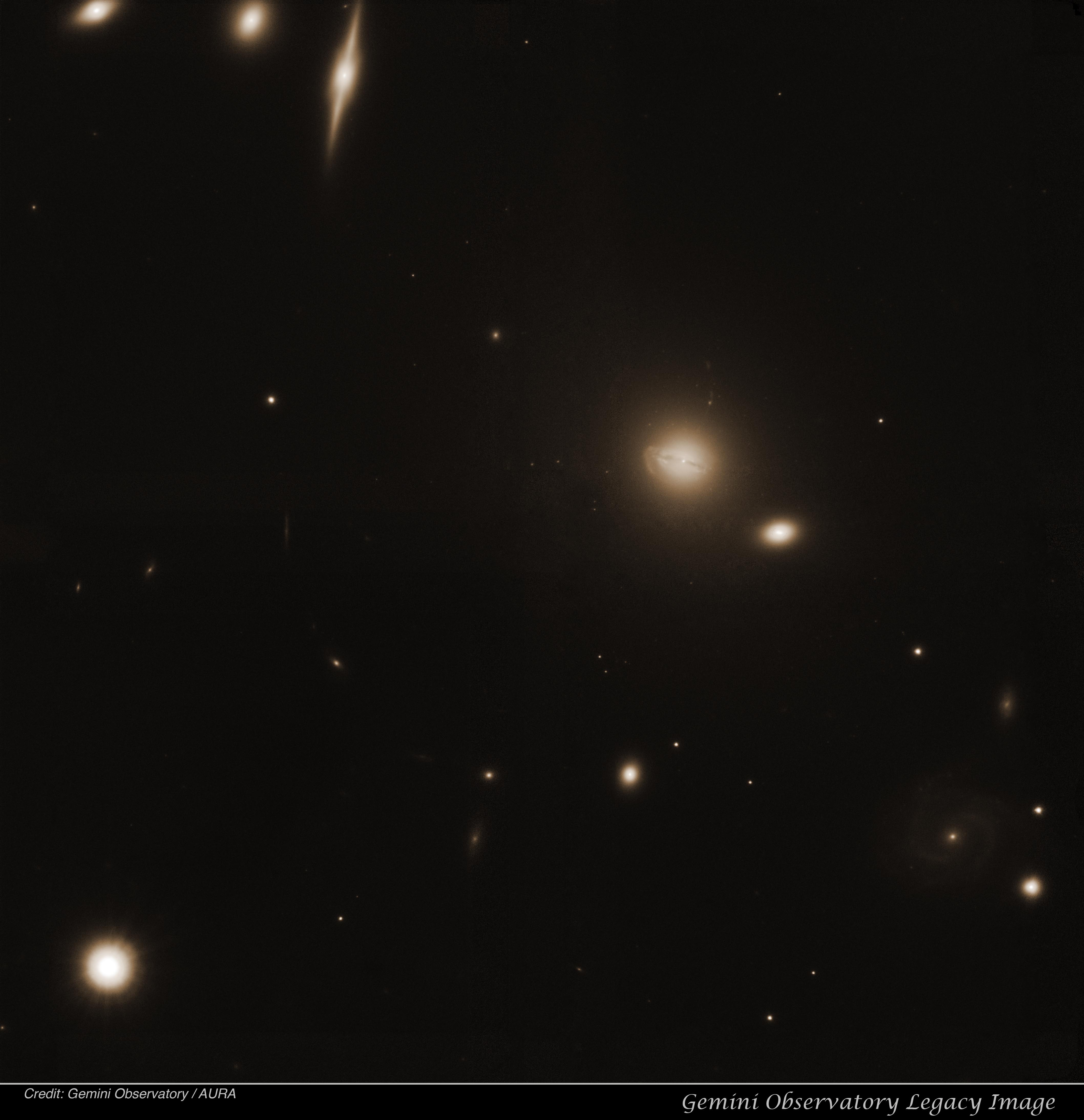} }}
\caption{Abell 780 - SV program led by Eleazar Rodrigo Carrasco. The image, made from a Ks filter has a field-of-view of 1.4 x 1.4 arcminutes and is oriented with north up. The total exposure time is 70 minutes.}
\label{fig:5}       % Give a unique label
\end{figure}

The second major reason why astronomers use MCAO is because of the enhanced sky-coverage. By sky-coverage, we mean the performance one can get at a given distance from the nearest Natural Guide Star (NGS). In a Single Conjugate AO (SCAO) system, even when using a LGS, the target of interest must lie close enough to the NGS used for Tip-Tilt measurements. GeMS is using three NGSs, 
% In English, acronyms can be plural, no issue with that, see for instance http://wps.ablongman.com/long_faigley_penguinhb_1/7/1977/506336.cw/
which can appear to be more restrictive than the SCAO-LGS mode, however, these NGSs can be anywhere in a 2 arcmin diameter acquisition field. Hence, the science target can be as distant as 60 arcsec from the NGS, and because of the MCAO correction, the performance will be essentially as good as if the target would lie closer to the NGS. This is particularly interesting in extra-galactic studies, that usually suffer from a low NGS and target density. As an example, Fig. \ref{fig:5} shows an image of Abell 780. Abell 780 (better known as Hydra A) is a rich cluster of galaxies 840 million light-years distant. For this target, only 2 NGS have been used, one of them lying on the bottom left of the image shown in Fig.\ref{fig:5}, the other one lying out of the field, on the top-left. Even with only 2 NGS, the performance is highly uniform over the field, with an average FWHM of 77 mas. Such performance, so far away from any usable NGS by a SCAO-LGS system, is unique to GeMS.

%n this field, the team led by Rodrigo Carrasco from Gemini wants to explore not only the structure of potential massive compact gal- axies in galaxy clusters but also the detailed properties of the massive galaxies with aver- age sizes. Looking for “signatures” that could be related to ongoing merger activity (such as tidal tails, clumps of star formation, etc.), they would be able to decide between dif- ferent competing evolutionary scenarios.

%extragalactic studies, only a few natural guide stars are available to provide AO corrections, which limits the areas of the sky available to study. With its five laser guide stars, GeMS in- creases the portion of the sky that can ben- efit from AO correction, and surpasses the previous generation of laser guide star AO systems. 

%Abell 780: This is a rich cluster of galaxies 840 million light-years distant. Better known as Hydra A, Abell 780 has been thoroughly studied at X-ray wavelengths, but its fine-scale structure has largely remained a mystery to astronomers at optical wavelengths. Recent studies, however, show the cluster to have a gravitationally bound structure of 27 galaxies, and a more strongly gravitationally bound structure of 14 galaxies. This GeMS/GSAOI image shows the cluster's core in unprecedented detail. Technical Data: 

\subsection{Astrometry with MCAO}

\begin{figure}[!ht]
\centerline{\resizebox{0.9\columnwidth}{!}{\includegraphics{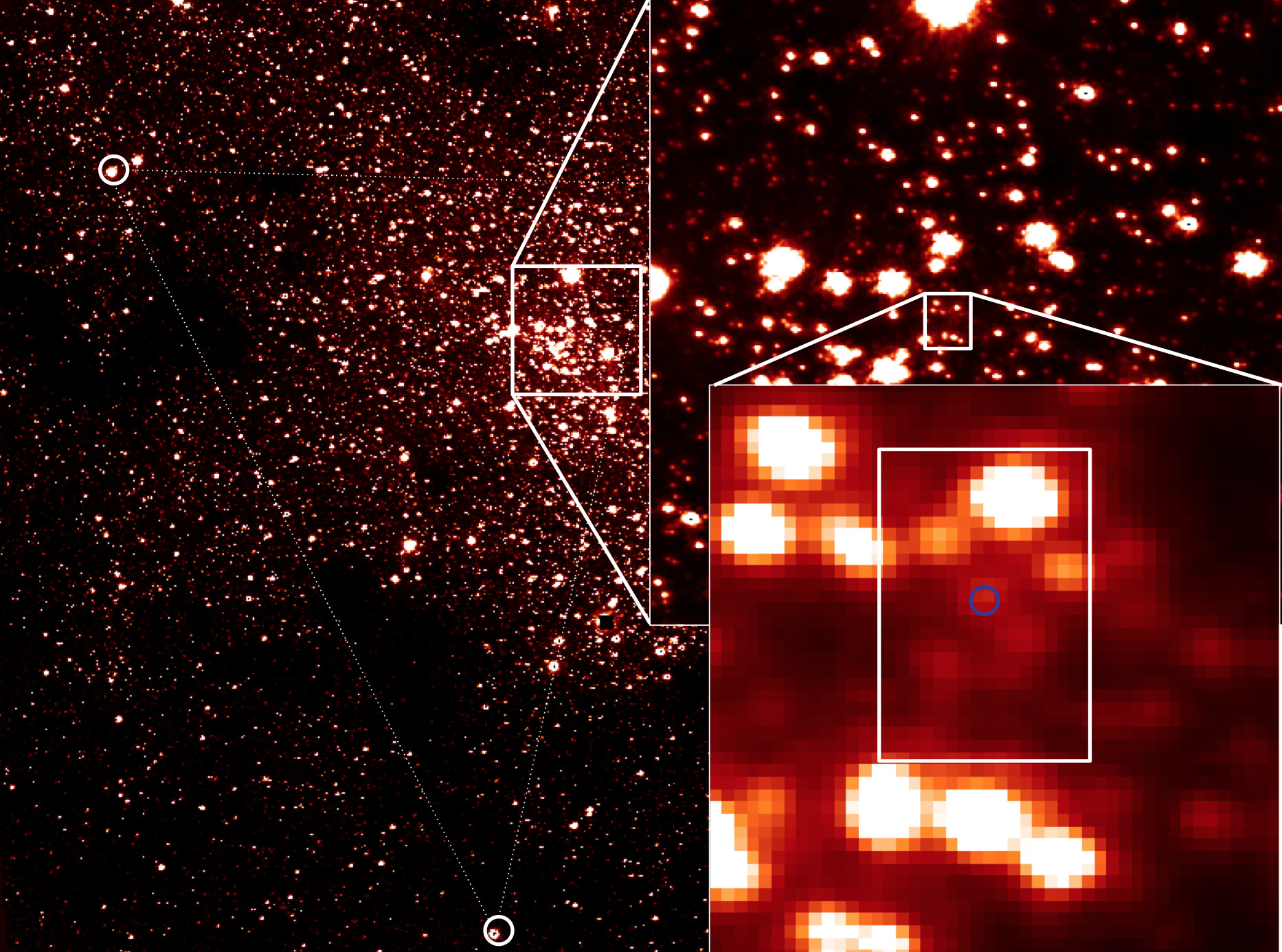} }}
\caption{The Galactic center at K band. }
\label{fig:6}       % Give a unique label
\end{figure}

Finally, another reason to use MCAO would be the potential of the system to do precision astrometry. Wide field adaptive optics systems have the potential to be a premier facility for precision astrometry due to the powerful combination of high spatial resolution and large field of view. In a single field, MCAO provides a large number of reference stars, and all of these references are of good quality, because of the uniform correction. Potential astrometric science cases cover a broad range of topics including exo-planets \cite{Ammons2013}, star formation, stellar evolution, star clusters, black holes and neutron stars, and the Galactic center. Many of these areas cannot be addressed with HST or GAIA due to insufficient resolution or lack of sensitivity at infrared wavelengths. As an example, the image of Fig. \ref{fig:6} is a view of the Galacti Center, around the location of the super massive black hole SgrA*. A high-resolution, multi-year observation of this region reveals the motion of the stars around the black hole, and allows a precise study of its physical characteristics \cite{eckart1999,ghez2000}. % Can't talk about that without citing both groups !
A first analysis of the GeMS/GSAOI astrometry performance has been presented in \cite{rigaut2012gems}. A more detailed analysis was presented at this conference by \cite{Lu2013}.

\section{The Future of GeMS}
It is expected that GeMS will be in regular operation basis (as a facility instrument) in November 2013. However, a number of upgrades are planned, especially for the most critical systems. These upgrades will be implemented as they become available. In particular, the emphasis will be given to increase the performance, and address the issues discussed in Sect. \ref{sec:1}. 

The first upgrade, planned for June/July 2014, is to restore a 3 DMs configuration, by installing back a DM at 4.5km. As described above, this will improve the performance and will make the system more robust to changes in turbulence conditions. The next upgrade, planned for the end of 2014, will the improvement in sensitivity of the NGSWFS. Due to design issues and alignment difficulties, the current NGSWFSs have a very low sensitivity, with a limiting magnitude around R=15.5. A project is currently under way at the Australian National University to build a replacement for the TT NGSWFSs, based on a single focal plane array covering the whole 2 arcmin field of view, reading out at 400Hz with less than 2 electrons read out noise. If the project is funded, it will boost significantly the TTWFS performance; nominally back to the expected level (limiting magnitude of 18.5). It will also drastically ease the acquisition procedure, and the need for lengthy probe mapping calibrations. Looking further away, an effort has been started to improve the optical beam quality of the 50W laser. The project, led by the Pontifica Universidad Cat{\'o}lica de Chile \cite{Bechet2013}, will implement a two-deformable mirror system in the BTO, in order to reduce the laser aberrations, and increase the LGS WFS measurements signal-to-noise ratio. 

Finally, the goal for the the next semesters will be to offer more science capabilities. Plans are to couple GeMS with FLAMINGOS-2, a near-infrared imager and multi-object spectrograph, as well as with GMOS the Gemini visible Multi-Object Spectrograph \cite{Hibbon2013}.

\acknowledgement
\noindent 
{\bf Acknowledgements}
\noindent
The authors would like to thanks all the PIs of the SV programs that send the images and the scientific rationale behind each program. These are John Bally, Robert Blum, E. Rodrigo Carrasco, Eiichi Egami, Alan McConnachie, Stuart Ryder, Lydia Stanghellini, Tim Davidge, David Flyod, Peter McGregor, Ronald Mennickent and Henri Plana. More images can be found on-line at http://www.gemini.edu/node/12020. 

% the followign is taken from the gemini website, on 2013 october 1rst:
\noindent
Based on observations obtained at the Gemini Observatory, which is operated by the 
Association of Universities for Research in Astronomy, Inc., under a cooperative agreement with the NSF on behalf of the Gemini partnership: the National Science Foundation (United States), the National Research Council (Canada), CONICYT (Chile), the Australian Research Council (Australia), Minist\'{e}rio da Ci\^{e}ncia, Tecnologia e Inova\c{c}\~{a}o (Brazil) and Ministerio de Ciencia, Tecnolog\'{i}a e Innovaci\'{o}n Productiva (Argentina).

\end{document}